\documentclass[apj]{emulateapj}
\usepackage{floatrow}
\usepackage{threeparttable}

\begin{document}

\slugcomment{To Appear in the Astrophysical Journal}

\title{A Low Frequency Survey of Giant Pulses from the Crab Pulsar}
\author{T. Eftekhari$^{1,2}$, K. Stovall$^{1}$, J. Dowell$^{1}$, F. K. Schinzel$^{1}$, G. B. Taylor$^{1}$ \bigskip}
\affil{1. Department of Physics and Astronomy, University of New Mexico, Albuquerque, NM 87106\\2. Harvard-Smithsonian Center for Astrophysics, Cambridge, MA 02138}

\email{tarraneh.eftekhari@cfa.harvard.edu}

\begin{abstract}

We present a large survey of giant pulses from the Crab Pulsar as observed with the first station of the Long Wavelength Array. Automated methods for detecting giant pulses at low frequencies where scattering becomes prevalent are also explored. More than 1400 pulses were detected across four frequency bands between 20 - 84 MHz over a seven month period beginning in 2013, with additional followup observations in late 2014 and early 2015. A handful of these pulses were detected simultaneously across all four frequency bands. We examine pulse characteristics, including pulse broadening and power law indices for amplitude distributions. We find that the flux density increases toward shorter wavelengths, consistent with a spectral turnover at 100 MHz. Our observations uniquely span multiple scattering epochs, manifesting as a notable trend in the number of detections per observation. These results are characteristic of the variable interface between the synchrotron nebula and the surrounding interstellar medium. 
\end{abstract}
\bigskip

\section{Introduction}
The average emission profile of the Crab Pulsar exhibits occasional bursts of increased intensity commonly referred to as giant pulses \citep{Staelin1968}. These individual pulses can exceed the average flux density by several orders of magnitude, becoming one of the brightest radio sources in the sky \citep{Jessner2005}. The short duration and high brightness temperatures of these bursts are indicative of non-thermal, coherent emission \citep{Hankins2003}. Nevertheless, the exact mechanisms responsible for these spurious bursts remain elusive. The majority of studies to date have focused on high time resolution radio observations at several GHz, where the effects of dispersion and scattering do not degrade the intrinsic nano to microsecond time resolution of detected pulses. More recently, several campaigns have emerged to study giant pulses and the effects of multi-path propagation at lower frequencies (e.g. \citet{Oronsaye2015}, at 193 MHz; \citet{Karuppusamy2012}, at 110-180 MHz; \citet{Bhat2007}, at 200 MHz).

In this paper, we present results from a low frequency survey of giant pulses from the Crab Pulsar as observed with the first station of the Long Wavelength Array (LWA1). In \S 2, we introduce the observations, followed by a discussion of pulse shapes at low frequencies in \S 3. The data reduction scheme and flux density calibration are discussed in \S 4 and \S 5. Finally, in \S 6, we examine pulse characteristics and discuss the implications of our results.

\section{Observations}
\subsection{The Long Wavelength Array}
The observations presented here were collected with the first station of the Long Wavelength Array (LWA1) radio telescope located in central New Mexico. LWA1 is co-located with the Karl G. Jansky Very Large Array and is comprised of 256 dual-polarized dipole antennas arranged over a 100 m x 110 m collecting area, with the longer elongation in the North-South direction allowing for the preservation of main-lobe symmetry at lower declinations. 

The LWA1 operates between 10 and 88 MHz. Individual dipole elements are digitized and combined to form four dual-polarized beams with independent pointings via a delay-sum beamforming technique. Each beam provides two frequency tunings -- configurable across the LWA1 frequency capability -- and 16 MHz of usable bandwidth per tuning. In addition to beamforming, LWA1 also has two transient buffer modes in which all dipoles observe the entire sky simultaneously. A second LWA station is currently undergoing commissioning and is expected to be online in early 2016. For an in-depth discussion of the design and science goals of LWA1, see \citet{Taylor2012} and \citet{Ellingson2013b}. 

The observations discussed here were collected over a period of seven months beginning in August 2013. Follow-up observations were obtained in October and November 2014 and in January 2015, for a total of 73 hours of observations, corresponding to roughly 37 TB of raw data. For each observation, two beams were centered on the Crab Pulsar (PSR B0531+21) at upper culmination, henceforth referred to as the Crab. As in \citet{Ellingson2013}, the center frequency tunings were 28 MHz, 44 MHz, 60 MHz, and 76 MHz, with 16 MHz of usable bandwidth, allowing for continuous coverage over the LWA1 frequency capability.  In all observations, the analog receivers (ARX) were configured to split bandwidth mode. As a result, signals below 30 MHz were attenuated, mitigating low frequency radio interference (RFI). To our knowledge, this sample represents the largest survey to date of giant pulses at these low frequencies.

\section{Giant Pulse Shape at Low Frequencies}

Due to changes in the electron density of the ISM, a given pulse emitted from the Crab will undergo multi-path propagation. A wave passing through a thin slab of electrons along the line of sight to the observer undergoes a phase change due to density fluctuations within the slab. These phase changes within the medium are greater at lower frequencies and contribute to a larger angular scatter for the emitted radiation \citep{Williamson1972}. An intrinsically narrow pulse is thus observed with some apparent broadening. This phenomenon is particularly prevalent at lower frequencies since the scattering dependence goes as $\nu^{-4}$ \citep{Lang1971}.  Pulse shapes at low frequencies are therefore unlike the intrinsically narrow pulses observed in the GHz regime and are instead characterized by a rapid rise followed by an exponential decay. Below 200 MHz, pulse shapes are best described by 

\begin{equation}
g(t) = t^{\beta}exp(-t/\tau_d)u(t)
\end{equation}

\noindent where $\tau_d$ is commonly referred to as the ``characteristic broadening time" of the pulse, $\beta$ describes the rise time of the leading edge, and $u(t)$ is the unit step function which takes on the value 1 for the duration of the pulse and is 0 otherwise \citep{Karuppusamy2012}; \citep{Ellingson2013}. The parameter $\tau_d$  has been known to fluctuate by factors of 2-10 over periods of days to months \citep{Rankin1973} Such variations are likely due to high density clouds within the nebula that pass along the line-of-sight. 

Further effects which may distort the observed data include synchrotron radiation from the galactic background. Such radiation has a steep frequency dependence ($\nu^{-2.6}$, \citep{Lawson1987}; \citep{Reich1988}) and appears as a contributing factor in the total system temperature. This occurrence is particularly noticeable for frequencies below 5 GHz. Similarly, the ionosphere can impart an additional time and phase delay which becomes a factor especially in interferometric arrays. Inhomogeneous clumps in the ionosphere contribute to differential delays which are difficult to account for and calibrate \citep{Stappers2011}.

\section{Data Reduction}

The first stage in the data reduction process is similar to that described in \citet{Stovall2015}. Initially, the raw, beamformed digital receiver (DRX) voltages were converted to the standard \texttt{PSRFITS} format \citep{Hotan2004} using the \texttt{writePsrfits2.py} utility from the LWA Software Library \citep{Dowell2012}. The data were then searched for RFI with \texttt{PRESTO}'s \texttt{rfifind} \citep{Ransom2001} using two second integration times, and the output mask was applied to the data for all subsequent processing. Because the Crab has been known to exhibit variability in its dispersion measure (DM) (as much as 0.01 pc $\mathrm{cm}^{-3}$ per month, see  \citet{Lyne1993}, the data were incoherently dedispersed into 200 DMs centered around the Crab's canonical DM (56.791 pc $cm^{-3}$, \citep{Counselman1971} in 0.01 step sizes. The specific parameters defined for incoherent dedispersion are outlined in Table 1. A dispersed pulse is shown in Figure 1. Following incoherent dedispersion, pulses were identified via a matched filter search using Eq. 1 with $\tau_d$ values of 10, 20, 30, 50, 100, and 200 time bins and a $\beta$ of 0.42 (varying $\beta$ seemed to have little to no effect on the identification of detected pulses). After identifying such a pulse, a fit to Eq. 1 was applied to the data and the resulting $\tau_d$ and $\beta$ values were recorded. Figure 2 depicts a giant pulse profile observed simultaneously in all four frequency bands.

\subsection{Identification of Giant Pulses}

Perhaps the biggest challenge in detecting giant pulses at low frequencies (where pulse broadening and RFI are ubiquitous) is inherent in developing automated detection methods. While pulse-matched filtering is commonly used in single pulse searches, it becomes increasingly more difficult at longer wavelengths where scattering within the host nebula and intervening ISM yield broad and inconsistent pulse shapes. As discussed above, ionospheric scintillation may result in amplitude modulations which obscure a pulse profile entirely. Ionized trails due to meteors are also offenders in matched filtering methods where radio emission results in dispersed signals on similar time scales \citep{Obenberger2014}. 

In light of these complications, pulse-matched filtering was utilized as only the first step in giant pulse detection. Template pulse shapes were cross-correlated at every DM. For a given pulse, the S/N vs DM was then plotted, where the S/N is defined as the area of the filter-matched pulse divided by the square root of the width. A gaussian-like peak centered about some DM confirmed detection of a pulse. This method is similar to comparing the S/N of a recorded pulse at zero DM (see \citet{Mickaliger2012}). Pulses with DMs bearing the highest S/N were stored as candidate pulses and used in subsequent pulse characterization. For all such pulses, the DM, S/N, best-fit $\tau_d$, and best-fit $\beta$ were recorded.

\section{Flux Density Calibration}

At the low frequencies and short baseline lengths of LWA1, absolute flux density calibration is particularly challenging. Diffuse radio continuum emission in the form of synchrotron radiation from the galactic background contributes to the total system temperature. The intensity of this galactic noise varies spatially across the sky and as a function of time over the course of the day. Beam sensitivity also varies as a function of pointing relative to zenith. Both the local sidereal time and the zenith angle therefore produce variations in the total system equivalent flux density (SEFD). In addition, ionospheric scintillation imparts stochastic fluctuations in the noise baseline which may corrupt SEFD measurements \citep{Ellingson2013b}. 

Measurements of the SEFD for LWA1 were obtained from over 400 hours of drift-scan observations of Cygnus A, Cassiopeia A, Taurus A, and Virgo A (see \citet{Schinzel2014}). In each case, a beam was fixed on upper culmination of the source where the observation began 1.5 hours prior to passage of the source through the center of the beam. The total peak power as the source transits through the beam is measured relative to the off-peak power measured when the source has not yet entered the side lobes. These results were used to determine the Stokes I SEFD for the telescope at zenith and across varying elevation angles. The SEFD was shown to remain fairly constant across much of the LWA1 frequency band, increasing only below 40 MHz \citep{Stovall2015}. For the pulses observed here, the following power ratio fraction is applied to the SEFD at zenith in order to obtain the system response for a particular elevation angle E in degrees:

\begin{equation}
P(E) = 166.625 \times E^{-1.251} + 0.401.
\end{equation}

In the case of the present work, our observations were limited to one hour in duration, and as such, variations in the SEFD due to shifting elevation angles are negligible. Flux densities presented here were obtained following improved calibration of LWA1 cable delays in March of 2013. It should be noted, however, that a total error of 50$\%$ is assumed for all LWA1 SEFD measurements: 25$\%$ error in the measurement of the SEFD at zenith and an additional 25$\%$ due to error in the fit to zenith angle. 

The nebular flux from the Crab also contributes to the total system temperature, although it is not the dominating factor as in most single dish telescopes in which the beam size is comparable to the size of the nebula. The nebula itself extends across a diameter of 5' \citep{Bietenholz1997} and is therefore unresolved by an LWA1 beam which is approximately 2$^{\circ}$ and 8$^{\circ}$ at 80 and 20 MHz, respectively \citep{Taylor2012}. For calculating the nebular flux density, we adopt the following equation:

\begin{equation}
\textrm S_{CN} = (1944 \textrm J\textrm y) (\nu/76 \textrm M\textrm H \textrm z)^{-0.27}
\end{equation}

\noindent where the spectral index $\alpha = 0.27$ (S$_{\nu} \propto \nu^{-\alpha}$) was first constrained by  \citet{Baars1977}. The flux factor was previously derived by \citet{Ellingson2013} via the extrapolation of values from separate measurements at 22.25 and 81.5 MHz (see \citet{Roger1969} and \citet{Parker1968}).

The SEFD and the flux density of the Crab Nebula were combined to obtain the total system noise, $S_{sys}$, and subsequently, the rms noise fluctuations in the time series, $\sigma$, given by $\sigma = S_{sys}/\sqrt{\Delta\nu\Delta t}$ \citep{McLaughlin2003}, where $\Delta\nu = 16$ MHz  and $\Delta t = 25$ ms. All pulses detected are based on a 4$\sigma$ threshold. 

\section{Results}

Results from each observing session are presented in Table 2. A total of 1458 pulses were detected over 73 hours of observations in our highest frequency band. Of these pulses, 506 were detected simultaneously in at least one other band. 143 pulses were detected simultaneously in three passbands, while only 8 pulses were confirmed as having been detected across all four frequency bands (a total of 33 pulses were detected at 28 MHz). The small number of detections at 28 MHz can likely be attributed to the ARX split-bandwidth configuration which was selected in order to suppress the effects of strong RFI below 30 MHz. In addition, high levels of scattering at the lowest frequencies lead to decreased detection rates. 

\subsection{Occurrence of Giant Radio Pulses}

The number of pulses per observation have been plotted over the initial 7 month period (see Figure 3). As evident from the plot, the number of pulses detected in an hour increases by roughly a factor of 2 over a period of three months, while the overall spread of detected pulses also increases slightly. Followup observations taken in late 2014 and early 2015 reveal that the detection rate drops back down, coinciding with initial rates. 

The apparent increase in the number of pulses over time cannot be attributed to instrumental effects such as gain variations. Drift scan observations of Cyg A indicate no evident changes in system sensitivity. Furthermore, the average flux density and dispersion measure over the same time period reveal no such trend. In addition, a preliminary analysis showed no correlation with gamma ray brightness as observed by Fermi. The average scattering timescales, however, are roughly a factor of 2 lower over the latter months relative to earlier months when the number of pulses detected are at a minimum (see Figure 4). Table 3 lists various parameters including the average DM, $\tau _d$, $\alpha$ (see \S 6.2), the average flux density, and the spectral index between 60 and 76 MHz for pulses from two distinct epochs of observations. The epochs are delineated via Figure 3, and refer to before and after the increase in number of pulses. Comparisons of $\tau_d$ between the two epochs reveal a decrease at both 60 and 76 MHz by the second epoch, although less drastic than the factor of two seen by binning all observable pulses (see Figure 4). These differences are characteristic of the increased spread in the number of detections at later times. 

These results may be indicative of the variable nature of the nebula which can lead to fluctuations in measured values of the broadening time, as discussed above. The longer the scattering time, the more the pulse is spread out, reducing the signal-to-noise such that weak pulses vanish below the noise floor. For a fixed pulse fluence, a narrow pulse corresponds to a higher peak flux, thereby increasing the likelihood of detecting the pulse. Subsequent pulses may also be lost among the scattering tails of their predecessors. The increase in scattering timescales are in particular thought to be associated with the interface between the synchrotron nebula and the surrounding medium  \citep{Hester1996}. Pressure discontinuities along this region manifest as enhancements in the scattering of radiation from the Crab. Thermal plasma structures in this region span a range of scale sizes which lead to variable scattering times as structures move in and out of the line-of-sight  \citep{Sallmen1999}.

\subsection{Characteristic Pulse Broadening Time}

Mean values for the characteristic broadening times for various studies have been plotted in Figure 5. Table 4 lists all values plotted. Broadening times for this study correspond to pulse-matched filters with the highest signal to noise. In the case of a Kolmogorov spectrum, the dependency is given by $\tau_d \propto \nu^{-4.4}$. A log-linear least squares fit to the data in Figure 4 (including values from this work) gives $\tau_d \propto \nu^{-3.5}$. If previous $\tau_d$ values as measured with LWA1 and presented in  \citet{Ellingson2013} are removed, the resulting power law dependence is given by $\tau_d \propto \nu^{-3.3}$. These dependencies are comparable to those reported by  \citet{Karuppusamy2012} ($\tau_d \propto \nu^{-3.2} \pm 0.1$), and suggest that previous LWA1 results span a separate scattering epoch. 

This deviation from the Kolmogorov spectrum reflects the inhomogeneous nature of the surrounding nebula, and in particular, implies that discrete filaments lead to perturbations of the frequency dependence of scattering  \citep{Cordes2001}. The variability in scattering times for our observations and the incongruity with earlier LWA1 results further suggest that a single power law dependence may not be sufficient for describing pulse broadening due to time variability of the scattering medium. Furthermore,  \citet{Ellingson2013} show a flattening in the frequency dependence beginning around 44 MHz whereas our data indicate no such trend. 

Power law fits were also made after binning the data into two segments (before and after the increase in pulses, as discussed above), and the results are listed in Table 3. Due to statistically insignificant sample sizes at the lower frequency tunings where the occurrence of pulses does not increase, the fit is made with only two data points at 60 and 76 MHz. In both cases, the slope is shallower than when all data points are used as in Figure 5. Most notably, however, is the move to a steeper slope in the second epoch, during which scattering times have decreased. 

\subsection{Pulse Amplitude Distributions}

Probability density functions (PDFs) for the flux densities at 76 and 60 MHz are shown in Figures 6 and 7 respectively. These distributions are well described by a power law. The slopes are obtained via maximum likelihood estimation, where the range over which the power law is applicable is determined by minimizing the Kolmogorov-Smirnov distance. Estimates for the slopes are obtained using the methods described in \citet{Alstott2014}. The slopes are given by $\alpha = $ 4.41 $\pm$ 0.09 (76 MHz) and $\alpha = $ 4.71 $\pm$ 0.17 (60 MHz). The brightest pulses detected correspond to peak fluxes of approximately 750 and 585 Jy at 76 and 60 MHz respectively. Below several hundred MHz, however, where the effects of scatter broadening produce exponential tails, measurements of the fluence provide a more accurate descriptor of the flux density and should correspondingly alter the power law dependencies.

For the eight pulses detected in all four frequency bands, a spectral index of +0.67 is given by the best fit. Pulses observed across the entire band represent an under-sampled population, and are likely not characteristic of the average giant pulse emission. The precise value of the spectral index derived here is as such not particularly meaningful. The positive slope, however, is consistent with a sharp spectral turnover at approximately 100 MHz  \citep{Rankin1970}, above which the flux density decreases more steeply with frequency \citep{Popov2007}. A power law fit to observations at 23 and 200 MHz by  \citet{Bhat2007} resulted in a spectral index given by +2.7.  \citet{Ellingson2013} suggest that extrapolations from  \citet{Karuppusamy2012} combined with initial LWA giant pulse results imply a spectral index below +2.7.

\section{Summary \& Discussion}

Over 1400 giant pulses from the Crab Pulsar have been detected with the LWA1 radio telescope in 73 hours of observations, compared to 33 pulses detected in 10 hours of observations as in  \citet{Ellingson2013}. This corresponds to an approximate increase in the rate of detection by a factor of six.  These differences may stem from improved calibration of the LWA1 cable delays since the initial giant pulse investigation. Additionally, results presented in  \citet{Ellingson2013} seem to suggest an altogether separate scattering epoch (see \S 6.2). Finally, detection methods presented here differ from those implemented in  \citet{Ellingson2013}, in which pulses were initially identified by eye. The use of pulse-matched filtering in the present work likely resulted in detections that would have gone otherwise unnoticed.

Our observations uniquely bracket a scattering epoch, given by the anticorrelation between average scattering timescales and the number of observable pulses. These fluctuating timescales represent the variable nature of the surrounding nebula, and in particular, provide an interesting probe of the nebula-ISM interface. 

A positive spectral index is obtained for those pulses observed concurrently in all four passbands. These results are consistent with a supposed spectral turnover at 100 MHz and indicate that giant pulse detection below 50 MHz becomes increasingly more difficult, relying on the brightest pulses which populate the tail end of the distribution. 

Continued observations of pulsars with the LWA1 are currently ongoing (see  \citet{Stovall2015}). Low frequency studies of pulsars are particularly well-suited for characterizing the effects of multi-path propagation through the interstellar medium. Such studies -- when combined with simultaneous observations at higher frequencies -- will allow for careful analysis of pulse morphologies across a range of frequencies, providing further constraints on the mechanisms responsible for pulsar emission. In particular, simultaneous observations of individual pulses spanning frequencies above and below the 100 MHz spectral turnover  will be particularly useful in characterizing the complex nature of giant pulse emission. 

\section*{Acknowledgements} Construction of the LWA has been supported by the Office of Naval Research under Contract N00014-07-C-0147 and by the Air Force Office of Scientific Research DURIP pro- gram. Support for operations and continuing development of the LWA1 is provided by the National Science Foundation under grants AST-1139963 and AST-1139974 of the University Radio Observatory program.

\newpage
\bibliographystyle{apj}
\bibliography{Eftekhari_T_2016}

\newpage

\begin{figure}[!hp]
\centering
\includegraphics[width=150mm]{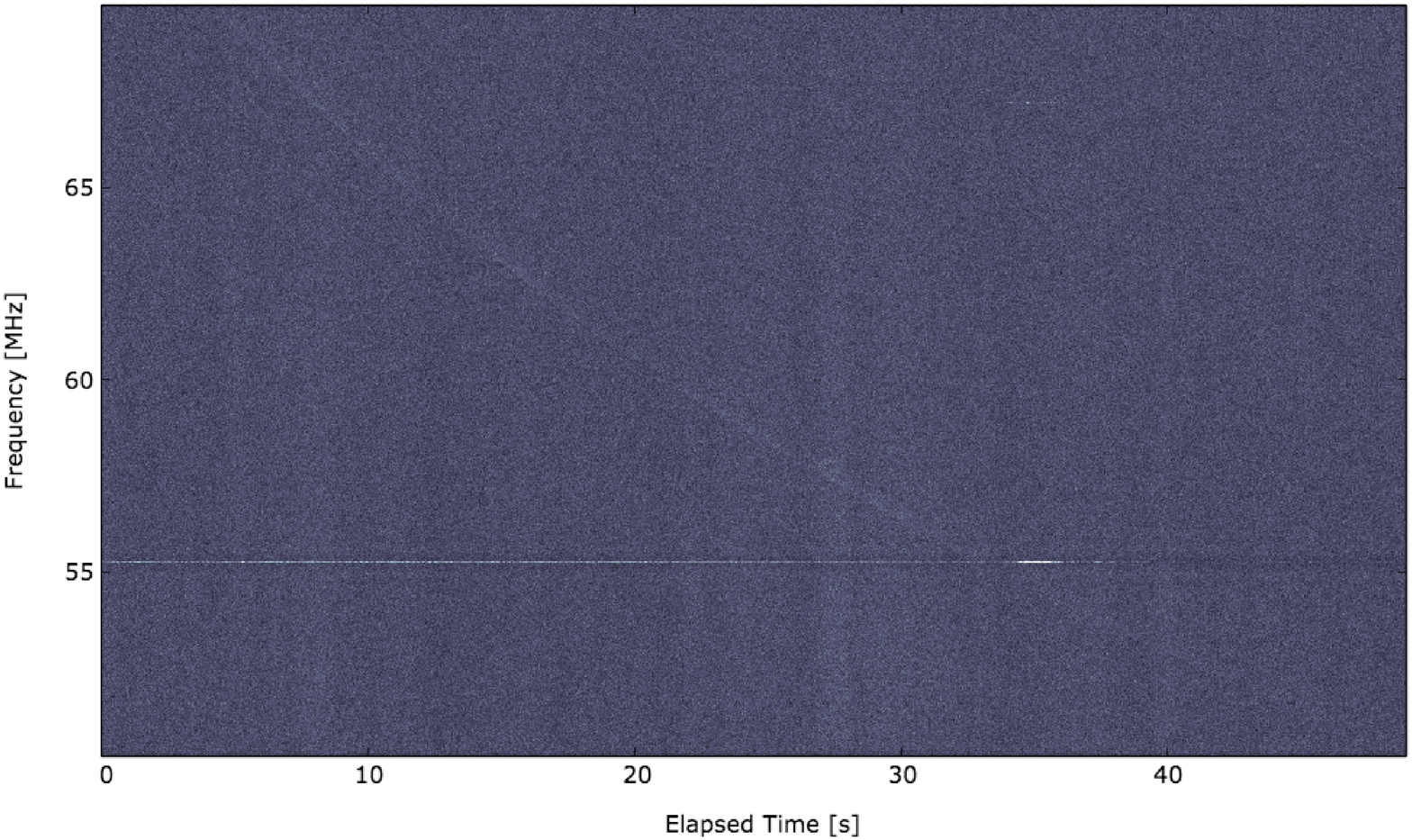}
\caption{Time averaged spectrum of a dispersed pulse in the 60 MHz band with 16 MHz of usable bandwidth and 0.01 second averaging.}
\end{figure}

\begin{figure}[!hp]
\centering
\includegraphics[width=100mm]{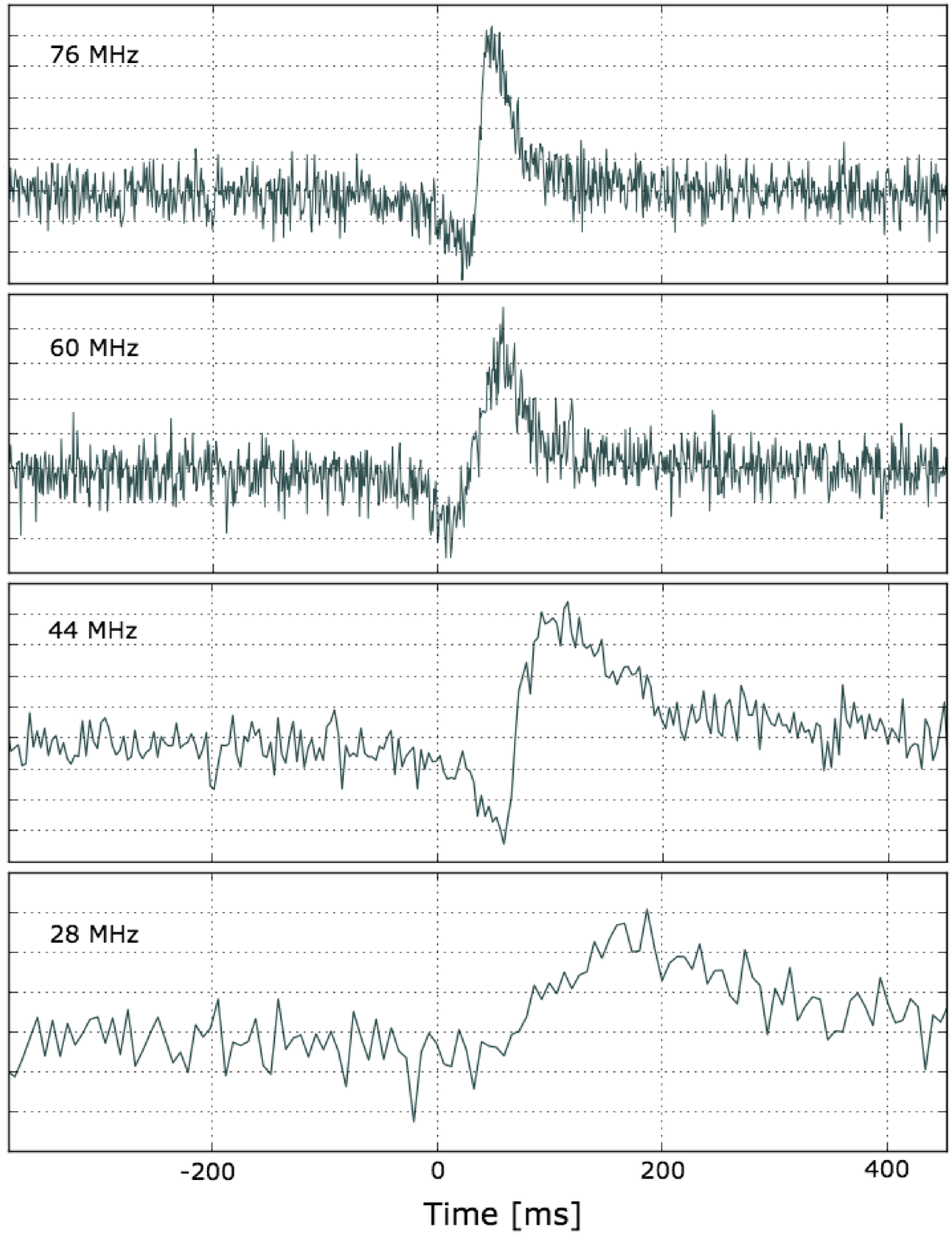}
\caption{A giant pulse detected in all four frequency bands. The amplitude dip before the pulse rise is an artifact of power spectrum whitening.}
\end{figure}

\begin{table}[hp]

\tiny
\centering
\begin{tabular}{c c c c c} 
\hline\hline 
Center Frequency [MHz]&Number of Channels&Channel Bandwidth [Hz]&Number of Bins&Bin Width [ms]\\
\hline 

76 & 16384 & 1196 & 4317184 & 0.8359\\
60 & 16384 & 1196 & 4317184 & 0.8359\\
44 & 65536 & 299 & 1077248 & 3.344\\ 
28 & 131072 & 150 & 536576 & 6.687\\

\end{tabular}
\label{table:nonlin}
\caption{Time and frequency resolutions utilized for incoherent dedispersion of pulses, as in \citet{Ellingson2013}.}
\end{table}

\begin{table}[hp]
	\tiny
	\centering 
	\begin{tabular}{c c c c c c c c}
	\hline\hline 
	MJD&28MHz&44MHz&60MHz&76MHz&4 Passbands&3 Passbands&2 Passbands\\ 
	\hline
	056529&0&2&10&30&0&2&7\\ 
	056532&0&2&-&20&-&0&2\\
	056538&0&0&5&24&0&0&4\\ 
	056545&0&4&8&14&0&2&4\\ 
	056556&2&4&9&16&0&0&0\\ 
	056559&1&4&12&18&1&2&5\\ 
	056573&0&5&7&12&0&3&3\\ 
	056580&0&6&6&11&0&2&1\\ 
	056588&2&4&11&10&0&0&7\\ 
	056589&1&1&10&15&0&5&6\\ 
	056590&-&6&9&12&-&1&6\\ 
	056595&0&4&3&9&0&3&3\\ 
	056596&-&-&9&9&-&-&6\\
	056597&0&6&9&13&0&0&2\\ 
	056598&1&5&11&12&0&0&4\\ 
	056601&0&5&1&4&1&3&2\\ 
	056603&1&8&3&4&0&0&1\\ 
	056604&0&4&11&14&0&0&9\\ 
	056605&0&7&9&16&0&2&6\\
	056607&-&-&7&10&-&-&3\\ 
	056608&0&2&13&18&0&2&6\\ 
	056610&0&5&7&7&0&0&4\\ 
	056611&1&12&9&10&0&5&3\\ 
	056612&0&3&3&6&0&0&1\\ 
	056615&0&6&14&10&0&3&6\\ 
	056618&2&4&12&17&1&1&9\\ 
	056619&0&10&12&17&0&6&3\\ 
	056622&-&2&9&9&-&2&4\\ 
	056624&0&10&-&-&-&-&0\\
	056625&-&-&18&33&-&-&16\\ 
	056627&-&-&11&14&-&-&6\\ 
	056628&-&6&7&13&-&3&3\\ 
	056630&2&8&14&19&2&3&7\\
	056632&3&5&10&11&2&2&4\\
	056636&1&3&-&-&-&-&0\\	
	056637&-&-&14&17&-&-&11\\ 
	056638&1&7&7&11&0&2&2\\ 
	056640&0&5&6&8&0&3&2\\ 
	056643&-&-&20&24&-&-&21\\ 
	056644&-&-&14&23&-&-&17\\ 
	056646&-&-&20&17&-&-&13\\ 
	056647&-&-&10&25&-&-&6\\ 
	056660&-&-&14&14&-&-&10\\
	056677&1&10&16&35&1&7&6\\ 
	056678&-&7&12&23&-&3&6\\ 
	056679&-&-&13&25&-&-&12\\ 
	056680&0&0&12&36&0&0&11\\ 
	056681&0&8&19&29&0&4&12\\ 
	056682&0&3&15&28&0&2&11\\ 
	056683&0&2&14&28&0&2&14\\ 
	056688&1&8&14&37&0&5&8\\ 
	056689&0&8&22&42&0&5&16\\ 
	056692&0&8&11&23&0&0&6\\ 
	056695&2&10&14&28&0&5&5\\ 
	056696&-&7&24&25&-&3&14\\
	056698&1&5&18&40&0&4&12\\
	056699&1&3&20&39&0&0&17\\ 
	056702&1&5&5&31&0&1&1\\ 
	056708&0&8&21&33&0&4&15\\ 
	056709&0&5&23&42&0&4&16\\ 
	056710&-&10&22&54&-&5&13\\ 
	056711&0&10&18&38&0&6&10\\ 
	056728&-&10&27&44&-&8&13\\ 
	056729&2&8&17&50&0&9&11\\ 
	056730&-&5&18&33&-&7&12\\ 
	056731&6&10&20&47&0&5&14\\ 
	056945&-&-&10&19&-&-&7\\	
	056946&-&-&8&17&-&-&3\\ 
	056974&-&-&3&22&-&-&2\\	
	057038&-&-&5&15&-&-&4\\
	\hline
	\end{tabular}
	\label{table:nonlin} 
\caption{Number of giant radio pulses detected per date in each frequency band. Also listed are the number of pulses detected simultaneously in multiple sidebands.}
\end{table}

\begin{figure}[!htb]
\centering
\includegraphics[width=110mm]{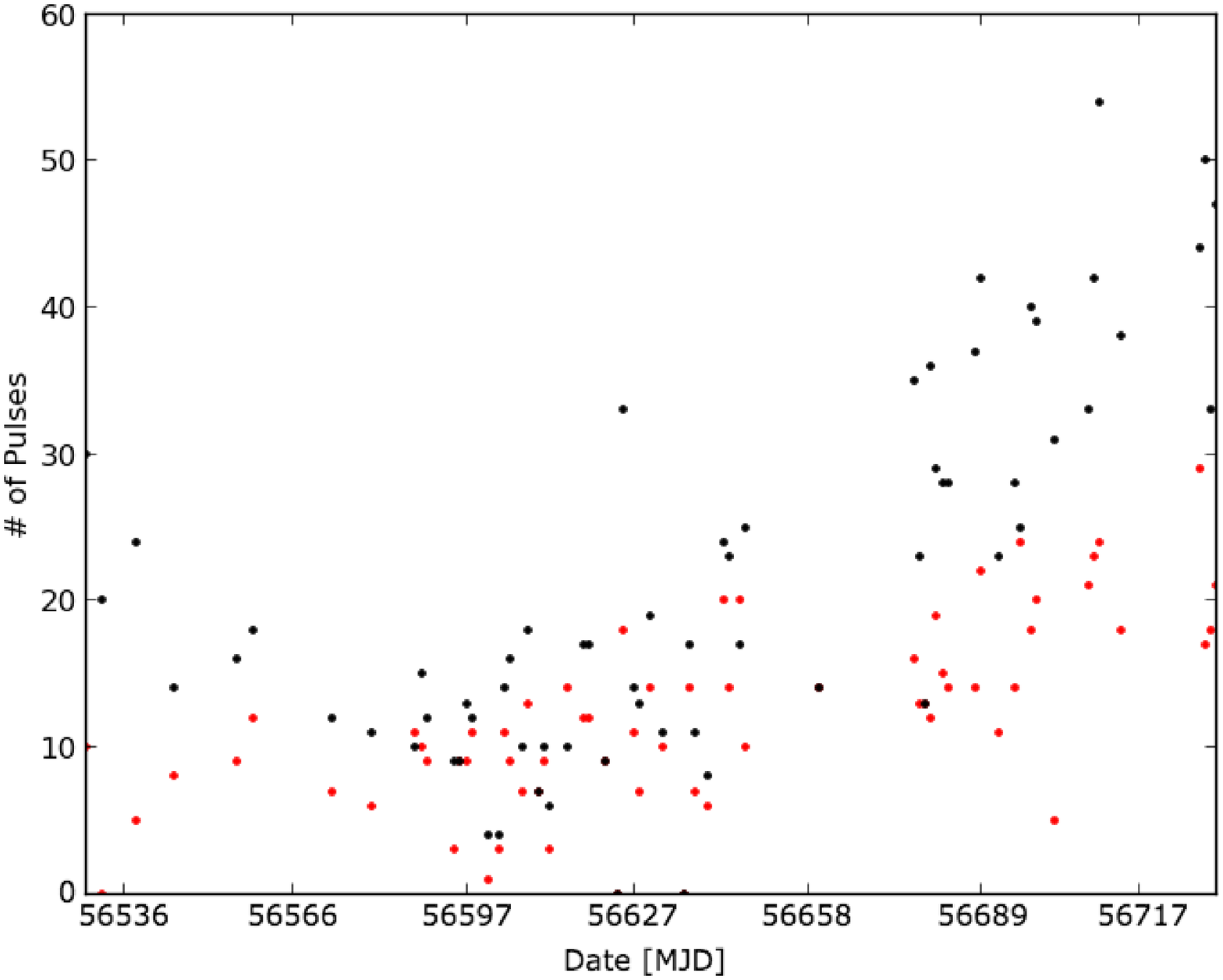}
\caption{Number of pulses detected per day at 76 MHz (black) and 60 MHz (red), with an increase  appearing between 56627 and 56720. Follow-up observations in 2015 have detection rates consistent with those seen in October - December 2013 (MJD 56566-56627).}
\end{figure}

\begin{figure}[!htb]
\centering
\includegraphics[width=110mm]{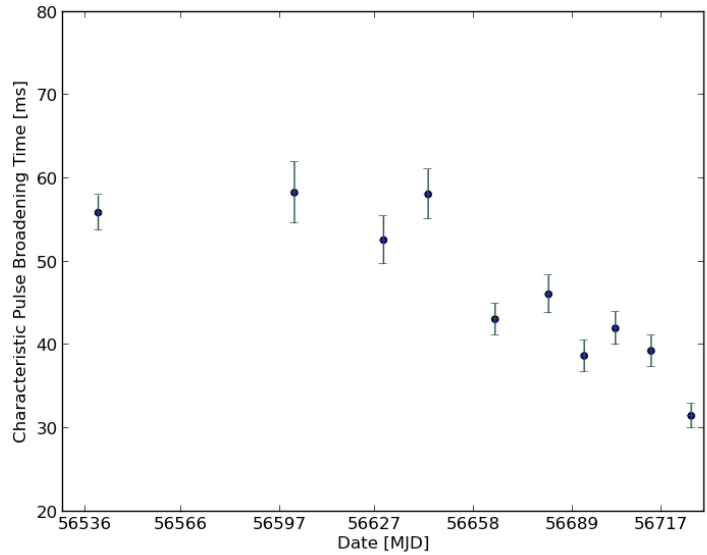}
\caption{Characteristic pulse broadening times $\tau_d$ binned and plotted as a function of time with error bars representing $\pm$1 $\sigma$ scatter within each bin.}
\end{figure}

\begin{table}[!htb]
\tiny
\begin{threeparttable}
\begin{tabular}{c c c c}
\hline\hline 
&Frequency [MHz] & Epoch I & Epoch II\\
\hline
DM$^\dagger$&76&56.89 $\pm$ 0.21&56.85 $\pm$ 0.14\\
DM&60&56.86 $\pm$ 0.19&56.87 $\pm$ 0.17\\
$\tau_{\rm d}^{*}$&76&54 $\pm$ 32&42 $\pm$ 23\\
$\tau_{\rm d}$&60&72 $\pm$ 50&71 $\pm$ 45\\
$\alpha$&&-1.22&-2.22\\
S$^{\ddagger}$&76&144 $\pm$ 68&144 $\pm$ 62\\
S&60&140 $\pm$ 57&143 $\pm$ 54\\
Spectral Index&&0.12&0.03\\
\hline
\end{tabular}
\begin{tablenotes}
%\multicolumn{4}{1} {$\dagger:$ Average DM in units of pc cm$^{-3}$.}\\
%\multicolumn{4}{1} {$*$: Scatter broadening time in ms. }\\
%\multicolumn{4}{1} {$\ddagger$: Flux density in Jy.}\\
\centering
\item $\dagger:$ Average DM in units of pc cm$^{-3}$.
\item $*$: Scatter broadening time in ms.
\item $\ddagger$: Flux density in Jy.
\end{tablenotes}
\end{threeparttable}
\label{table:nonlin} 
\caption{Parameters as calculated for separate scattering epochs where I and II refer to first and second epochs respectively.}
\end{table}

\begin{figure}[!htb]
\centering
\includegraphics[width=110mm]{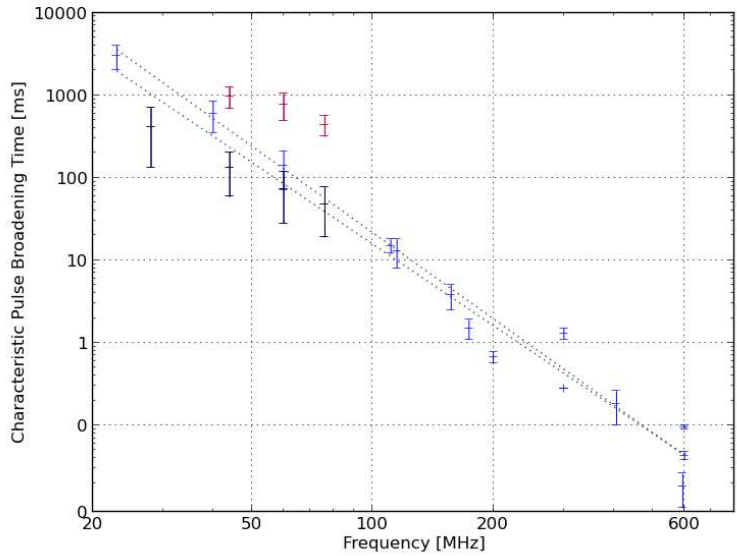}
\caption{Characteristic pulse broadening times $\tau_d$ for giant pulses detected in this work (black), a previous LWA1 campaign (red, \citet{Ellingson2013}), and previous studies as listed in Table 4 (blue). The error bars signify $\pm 1 \sigma$ from the mean. A log-linear least squares fit to all of the data points (top line) resulted in $\alpha$ = -3.5. The lower line depicts a fit to all data barring previous LWA1 measurements ($\alpha$ = -3.3).}
\end{figure}

\begin{table}[!htb]
\small
\centering
\begin{tabular}{c c c}
\hline\hline 
$\nu$ [MHz]  & $\tau_d$ [ms]  & Reference\\
\hline 

23& 300 $\pm$ 1000& \citet{Popov2006}\\ 
28& 417 $\pm$ 284& $\textit{This work}$\\
40& 132 $\pm$ 73& $\textit{This work}$\\
40& 600 $\pm$ 250& \citet{Kuzmin2002}\\ 
44& 978 $\pm$ 287& \citet{Ellingson2013}\\ 
60& 140 $\pm$ 70& \citet{Kuzmin2002}\\ 
60& 768 $\pm$ 273& \citet{Ellingson2013}\\ 
60& 73 $\pm$ 45 & $\textit{This work}$ \\ 
76& 439 $\pm$ 122& \citet{Ellingson2013}\\ 
76& 48 $\pm$ 29& $\textit{This work}$ \\ 
111& 15 $\pm$ 3 & \citet{Popov2006}\\ 
115& 13 $\pm$ 5 & \citet{Staelin1970}\\ 
157& 3.8 $\pm$ 1.3 & \citet{Staelin1970}\\ 
174& 1.5 $\pm$ 0.4 & \citet{Karuppusamy2012} \\ 
200& 0.670 $\pm$ 0.100 & \citet{Bhat2007}\\ 
300& 1.3 $\pm$ 0.2 & \citet{Sallmen1999} \\ 
300& 0.28 & \citet{Sallmen1999} \\\ 
406& 0.18 $\pm$ 0.08 & \citet{Kuzmin2002}\\ 
594& 0.018 $\pm$ 0.008 & \citet{Kuzmin2002}\\ 
600& 0.095 $\pm$ 0.005 & \citet{Sallmen1999} \\\ 
600& 0.043 $\pm$ 0.005 & \citet{Popov2006}\\ 

\end{tabular}
\label{table:nonlin} 
\caption{Characteristic pulse broadening times $\tau_d$ from various measurements as in \citet{Ellingson2013}, modified to include measurements from this work. }
\end{table}

\begin{figure}[!htb]
\centering
\includegraphics[width=110mm]{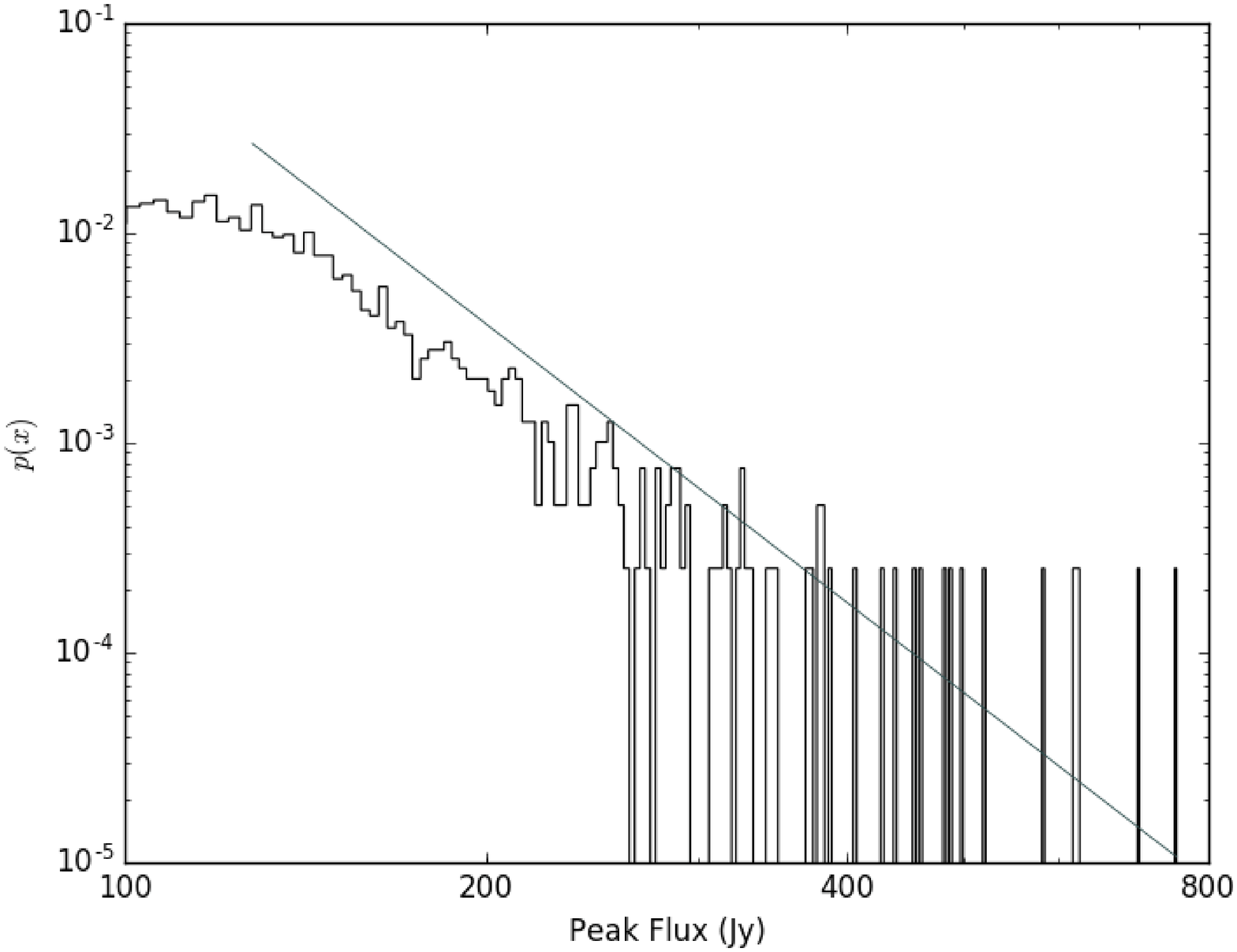}
\caption{PDF of peak flux densities at 76 MHz with a power law probability distribution corresponding to $\alpha = $ 4.41 $\pm$ 0.09. The brightest giant pulse detected at 76 MHz corresponds to a peak flux of $\sim$750 Jy.}
\end{figure}

\begin{figure}[!htb]
\centering
\includegraphics[width=110mm]{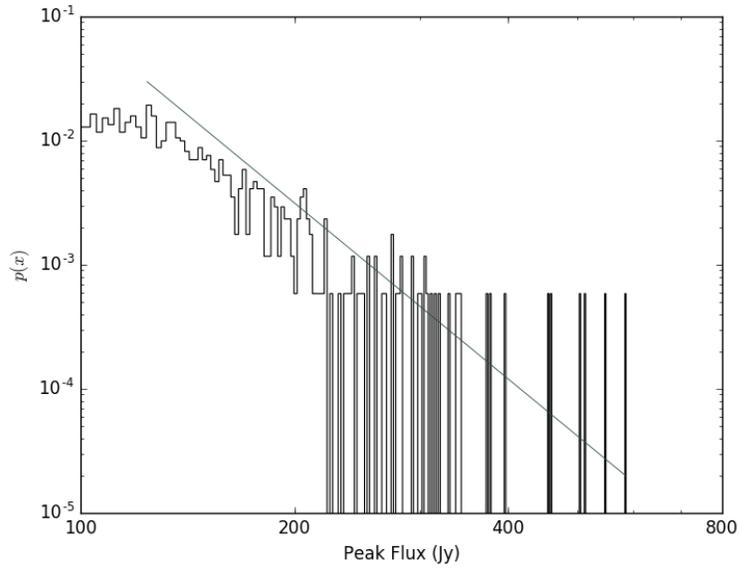}
\caption{PDF of flux densities at 60 MHz. The power law probability distribution is given by $\alpha = $ 4.71 $\pm$ 0.17. The brightest pulse detected at 60 MHz corresponds to a peak flux of  $\sim$585 Jy.}
\end{figure}

\end{document}